**Non-piezoelectric effects in piezoresponse force microscopy**

Daehee Seol[1], Bora Kim[1] and Yunseok Kim[1*]

[1]*School of Advanced Materials Science and Engineering, Sungkyunkwan University (SKKU), Suwon, 440-746, Republic of Korea*

[*] Address correspondence to yunseokkim@skku.edu

**Abstract**

Piezoresponse force microscopy (PFM) has been used extensively for exploring nanoscale ferro/piezoelectric phenomena over the past two decades. The imaging mechanism of PFM is based on the detection of the electromechanical (EM) response induced by the inverse piezoelectric effect through the cantilever dynamics of an atomic force microscopy. However, several non-piezoelectric effects can induce additional contributions to the EM response, which often lead to a misinterpretation of the measured PFM response. This review aims to summarize the non-piezoelectric origins of the EM response that impair the interpretation of PFM measurements. We primarily discuss two major non-piezoelectric origins, namely, the electrostatic effect and electrochemical strain. Several approaches for differentiating the ferroelectric contribution from the EM response are also discussed. The review suggests a fundamental guideline for the proper utilization of the PFM technique, as well as for achieving a reasonable interpretation of observed PFM responses.

# 1. Introduction

The increasing demand for miniaturized electronic devices has prompted the development of novel techniques for characterizing material properties accurately at the nanoscale.[1, 2] In this perspective, scanning probe microscopy (SPM) has enabled new approaches for evaluating nanoscale material properties. Variants of SPM have been developed and extensively utilized to explore elastic,[3-5] electrical,[6-9] electrochemical,[10-12] magnetic,[13-16] and various functional properties[17-20] in the last few decades. Among them, piezoresponse force microscopy (PFM) has been particularly useful for exploring ferro/piezoelectric properties, owing to its remarkably high resolution and ease of use.[21-24] To date, the capability of PFM for probing nanoscale ferro/piezoelectric phenomena has been extensively demonstrated in ferroelectric oxide systems.[25] The scope of PFM applications has recently expanded into the fields of polymeric,[26-28] biological,[29-31] and organic–inorganic hybrid materials,[32-35] and also to two-dimensional materials,[36-38] with a view to investigating ferro/piezoelectricity. However, owing to the intrinsic imaging mechanism of PFM, several artifacts related with topographical features[39, 40] and background signals[41-43] can affect PFM measurements. Apart from these artifacts, there are concurrent non-piezoelectric effects, which contribute directly to the PFM response. In other words, the non-piezoelectric effects can cause an additional electromehcanical (EM) response that can be simultaneously detected as a PFM response and thereby distort the PFM measurement. This problem strongly motivates an understanding of the operational principle underlying PFM and of non-piezoelectric effects, to explicitly interpret PFM measurements.

## 1.1. Principle of PFM

Ferroelectric materials display a spontaneous polarization that can be switched by the application of an external electric field. They also possess piezoelectric behavior,[44, 45] described as a reversible linear relation between a mechanical deformation and an electric field. In general, the process of mechanical deformation, *i.e.* surface volume change (an expansion or contraction), induced by an electric field is referred to as the inverse piezoelectric effect in a ferro/piezoelectric material. In PFM, a periodic surface volume change can be induced by applying an ac voltage to the conductive tip of an atomic force microscope (AFM), then detecting the EM response through the AFM cantilever.[46] This particular EM response constitutes the PFM response and is quantified in terms of its amplitude and phase, by comparing the input ac voltage with periodic surface vibrations (the output signal from the sample) using a lock-in amplifier, as outlined in Fig. 1(a).[21] In general, the PFM amplitude ($A$) and phase ($\varphi$) signals characterize the magnitude of the piezoelectricity and the polarization direction, respectively. Together, they make up the piezoresponse $A \sin(\varphi)$. Analyzing the vertical (lateral) cantilever deflections, as shown in Fig. 1(b) (Fig. 1(c)), yields information on the vertical (lateral) component of the polarization, an approach referred to as vertical (lateral) PFM.[47, 48] On the basis of the operational mechanism, vertical and lateral PFMs can provide profound insight into nanoscale ferro/piezoelectric phenomena such as piezoelectricity,[49-52] polarization switching dynamics,[53-56] domain growth, and wall motions.[57-60] Three-dimensional domain structures can also be explored by combining vertical and lateral PFMs.[61-65]

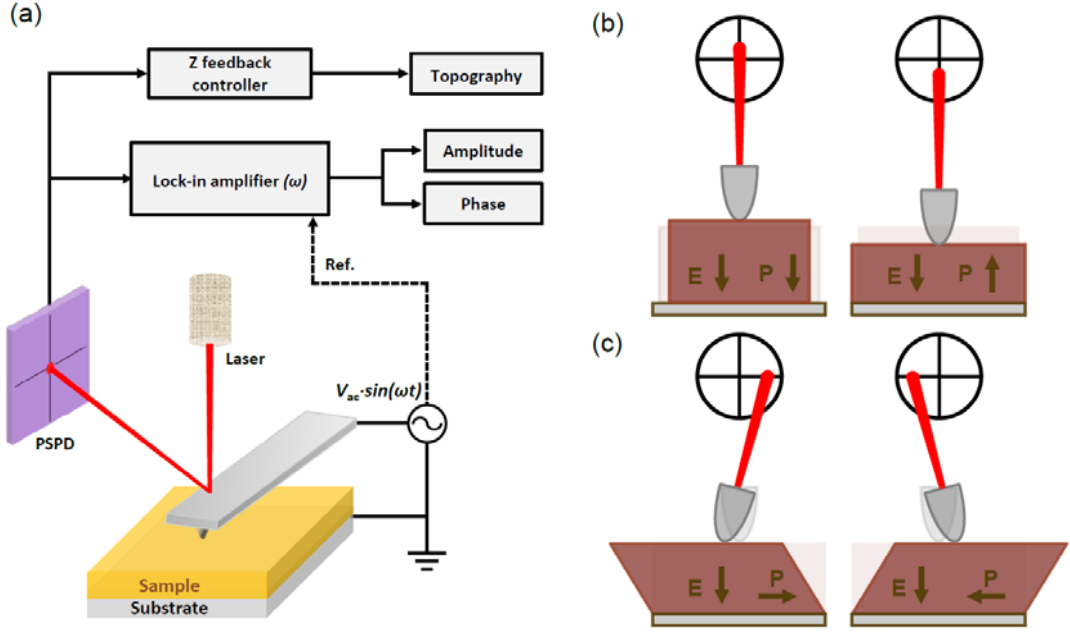

**Fig. 1.** (a) Schematic PFM setup, and the principle of (b) vertical and (c) lateral PFMs

The observed PFM response is generally expressed as

$$(PR)_\omega = d_{eff} V_{ac} \sin(\omega t) \tag{1}$$

where $d_{eff}$, $V_{ac}$, $\omega$, and $t$ are the effective piezoelectric coefficient, magnitude and frequency of the ac voltage, and time, respectively. The observed response by PFM is thus directly governed by piezoelectric properties such as the piezoelectric tensor of ferro/piezoelectric materials. A conventional PFM measurement is performed by applying a single ac frequency that is generally far from the contact-resonance frequency of the cantilever. However, resonance-enhancement techniques based on the use of the contact resonance frequency have been suggested. These techniques, which include dual ac resonance tracking (DART)[66, 67] and band excitation (BE),[68, 69] significantly enhance the signal-to-noise ratio of the observed PFM response, which facilitates the achievement of vertical (z-height) resolutions at the picometer level. Such techniques have been used extensively for PFM imaging and spectroscopic approaches, *e.g.*, switching spectroscopy PFM,[70] to explore local ferro/piezoelectric properties at the nanoscale.[71, 72]

## 1.2. Origins of the EM response in PFM

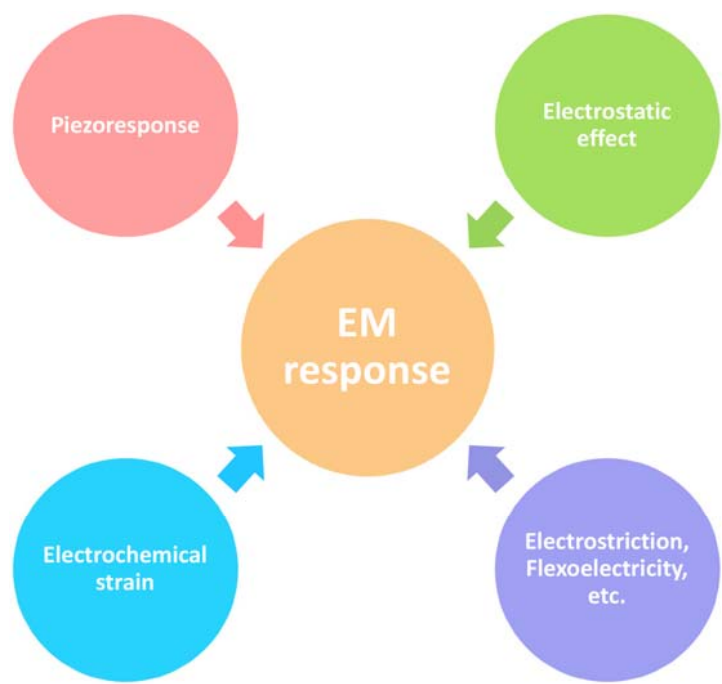

**Fig. 2.** Schematic diagram of origins that can induce EM response.

Although the piezoresponse of ferro/piezoelectric materials is of interest as the main origin of the induced EM response in the PFM, several other non-piezoelectric effects can also contribute to the EM response, as illustrated in Fig. 2. These non-piezoelectric effects not only hinder accurate PFM measurements but also, in some cases, cumulatively dominate the measured

PFM response.[73-75] It is therefore very important to acknowledge and understand their influence to allow an accurate interpretation of material properties measured by PFM.

This review primarily discusses two major non-piezoelectric effects that directly affect the interpretation of PFM results: the electrostatic effect and electrochemical strain. Section 2 addresses their definition and influence on the PFM response, and briefly introduces some relatively minor non-piezoelectric effects. Section 3 discusses recently reported AFM-based methods for differentiating ferroelectric contribution from the EM response.

## 2. Non-piezoelectric contributions to the PFM response

### 2.1. Electrostatic effect

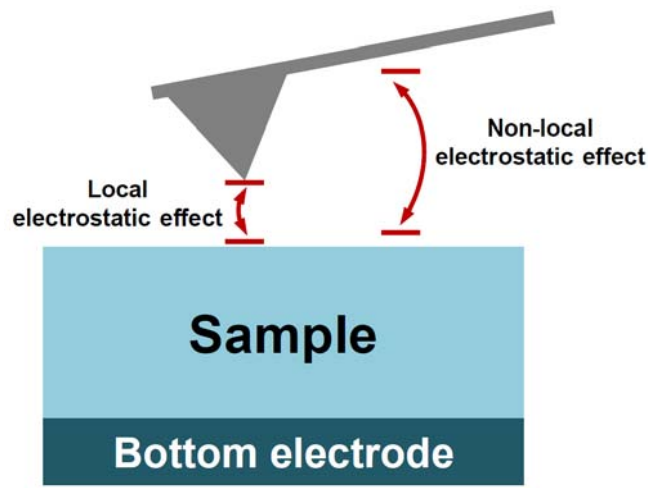

**Fig. 3.** Local and non-local electrostatic effects in the AFM system

The electrostatic effect is a parasitic interaction between the AFM tip/cantilever and the sample surface, regardless of the operational modes (*i.e.*, contact or non-contact modes). The electrostatic effect originates essentially from the Coulombic electrostatic force between the

AFM tip/cantilever and the sample surface,[75, 76] described as[77, 78]

$$F_{ES} = \frac{1}{2}\frac{dC}{dz}V^2. \qquad (2)$$

where *k*, *C*, and *z* are the cantilever spring constant, capacitance, and tip-sample distance, respectively. In this case, the applied ac and dc voltages give a total voltage $V = V_{dc} + V_{ac}sin(\omega t)$, and give rise to an additional surface displacement

$$x_{ES} = \frac{F_{ES}}{k} = \frac{1}{2k}\frac{dC}{dz}(V_{dc} + V_{ac}\sin(\omega t))^2. \qquad (3)$$

Thus, the first-harmonic components of the cantilever deflection induced by the electrostatic force can be derived as

$$(PR)_{\omega,ES} = k^{-1}C'V_{dc}V_{ac}\sin(\omega t). \qquad (4)$$

Furthermore, the influence of the additional contribution from the contact-potential difference (CPD) between the AFM tip/cantilever and the sample surface ($V = V_{dc} + V_{CPD} + V_{ac}\sin(\omega t)$) can also be included:[79]

$$(PR)_{\omega,ES} = k^{-1}C'(V_{dc} + V_{CPD})V_{ac}\sin(\omega t). \qquad (5)$$

This electrostatic effect can occur between the AFM tip and the sample surface (a local electrostatic effect) as well as between the AFM cantilever and the sample surface (a non-local electrostatic effect), as shown in Fig. 3. Such local and non-local electrostatic effects can arise from the various electrical interactions between the AFM tip/cantilever and sample surface, such as the capacitive interaction,[80, 81] the CPD,[79, 82] the application of an external voltage to the entire sample, and the corresponding injected charges on the sample surface.[83-85]

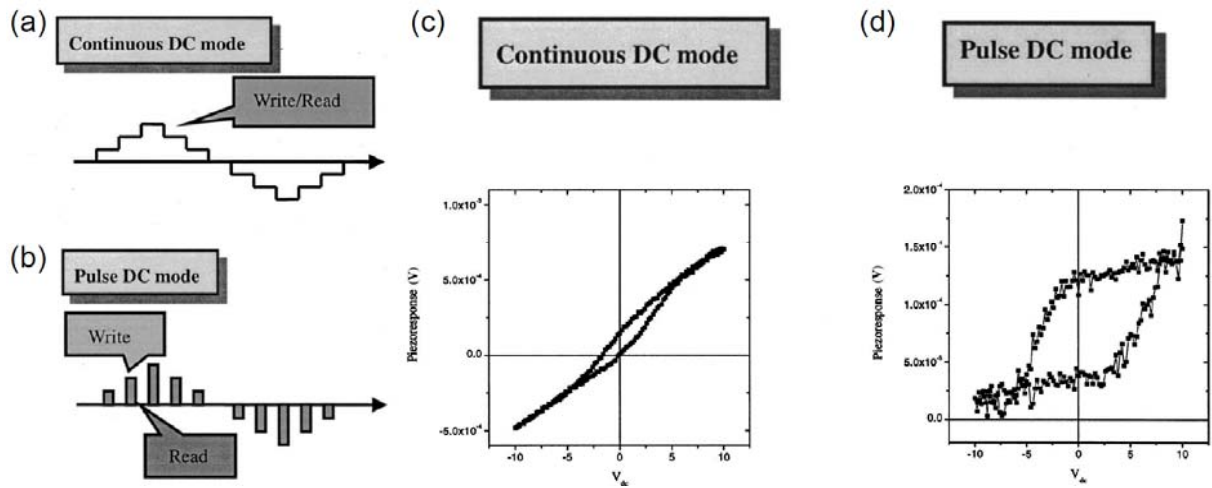

**Fig. 4.** Schematic diagrams of (a) continuous dc and (b) pulsed dc modes. Piezoresponse hysteresis loops obtained by (c) continuous dc and (d) pulsed dc modes in a Pb(Zr$_{0.4}$Ti$_{0.6}$)O$_3$ thin film. Reproduced with permission from [S. Hong, J. Woo, H. Shin, J. U. Jeon, Y. E. Pak, E. L. Colla, N. Setter, E. Kim and K. No, J. Appl. Phys. 89 (2001) 1377]. Copyright 2001, AIP Publishing LLC [81].

Hong *et al.* reported a significant influence of the electrostatic effect on piezoresponse hysteresis loop measurements.[81] A piezoresponse hysteresis loop is generally measured using a triangular waveform, which can be of two kinds, referred to as continuous and pulsed dc modes, as shown in Figs. 4(a) and (b).[81] In the continuous dc mode, the "read" and "write" operations are performed simultaneously by the dc and ac components, respectively, whereas they are performed separately in the pulsed dc mode. In other words, the "read" operation is performed at zero dc voltage (off-field) in the pulsed dc mode.

Figures 4(c) and (d) show the piezoresponse hysteresis loops obtained from a Pb(Zr$_{0.4}$Ti$_{0.6}$)O$_3$ (PZT) thin film using the continuous and pulsed dc modes, respectively. The hysteresis loop measured in the continuous dc mode appear distorted and slanted, compared to that obtained in the pulsed dc mode. The strong electrostatic effect induced by the applied dc voltage in the on-field state can distort the shape of the piezoresponse hysteresis loop and make it slanted and thin. This can give rise to misinterpretations of ferro/piezoelectric properties of the measured material, suggesting incorrect values for the magnitude of the piezoresponse and for the coercive voltage of polarization switching. Thus, appropriate piezoresponse hysteresis loop can be acquired in pulsed dc mode, by reducing the electrostatic effect.

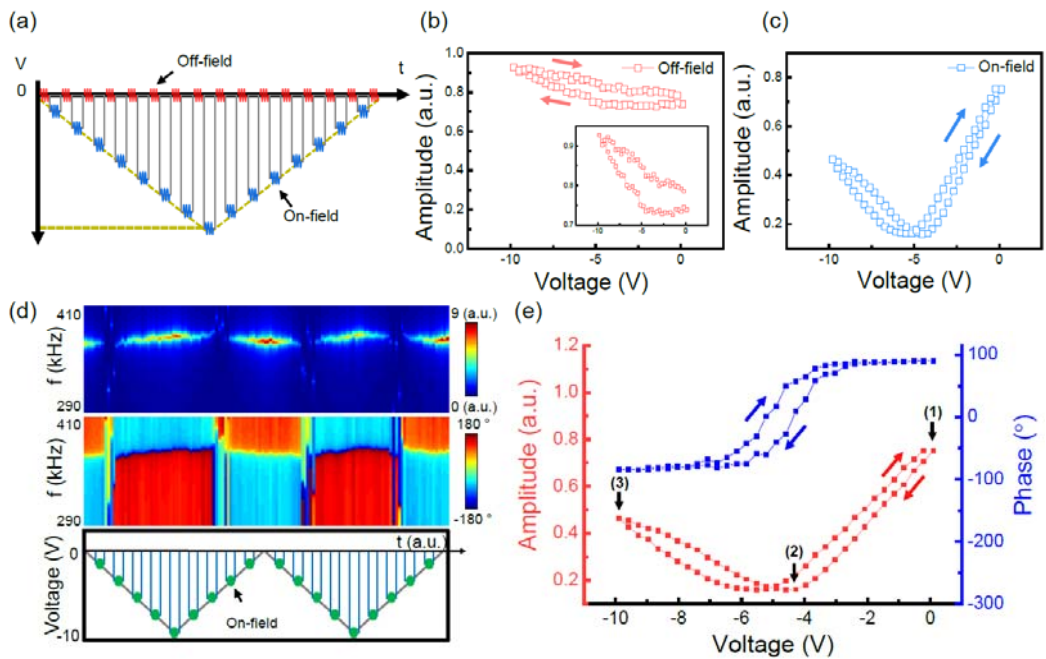

**Fig. 5.** (a) Unipolar negative voltage waveform and the corresponding (b) off- and (c) on-field unipolar amplitude hysteresis loops. The inset in (b) shows the hysteresis loop magnified. (d) (Bottom panel) Applied voltage waveform and the corresponding (top) PFM amplitude and (middle) phase spectra in the on-field state. (e) On-field unipolar amplitude and phase hysteresis loops. Reproduced with permission from [B. Kim, D. Seol, S. Lee, H. N. Lee, and Y. Kim, Appl. Phys. Lett. 109 (2016) 102901]. Copyright 2016, AIP Publishing LLC [84].

Despite thorough investigations of the influence of the electrostatic effect on the piezoresponse hysteresis loop, this influence was observed mainly under conditions where polarization switching occurred. Under such circumstances, the electrostatic effect cannot be clearly investigated because of interference from the polarization switching. Kim *et al.* reported the influence of the electrostatic effect on the hysteresis loop in the absence of polarization switching by using a unipolar voltage sweep on an epitaxial PZT thin film, as depicted in Fig. 5.[84] In this case, since the film had a uniform upward polarization, a negative voltage was applied to avoid polarization switching. Figures 5(b) and (c) show off- and on-field unipolar amplitude hysteresis loops, respectively. The off-field loop in Fig. 5(b) displays weak hysteresis in the shape of a half butterfly. This hysteric behavior may originate from a local electrostatic effect due to charge injection because the upward polarization is not changed by the negative voltage. Intriguingly, the unipolar amplitude hysteresis loop in the on-field state displays a butterfly-like shape despite the absence of polarization switching. This clear ferroelectric-like behavior is further apparent in the amplitude and phase spectra in Fig. 5(d). This behavior is related to the electromechanical neutralization between the piezoresponse of the polarization and the electrostatic effect, because the results were observed in the on-field state. Note that, electromechanical neutralization indicates the interaction between the piezoresponse of the polarization and the applied dc voltage during the on-field hysteresis loop measurement. Because the piezoresponse contributes fully to the hysteresis loop response at the initial point (0, $V_{dc}$), labeled (1) in Fig. 5(e), the decremental amplitude may be correlated with the electromechanical neutralization on the ferroelectric surface. After the first inflection point, labeled (2), the increase of the measured PFM amplitude indicates that the electrostatic effect comes to dominate the piezoresponse of the polarization, and hence the ferroelectric-like hysteresis loop can develop. The observed phenomena thus demonstrate that the ferroelectric-like hysteresis loop can be generated by the electrostatic effect through the interaction with the piezoresponse of the polarization.

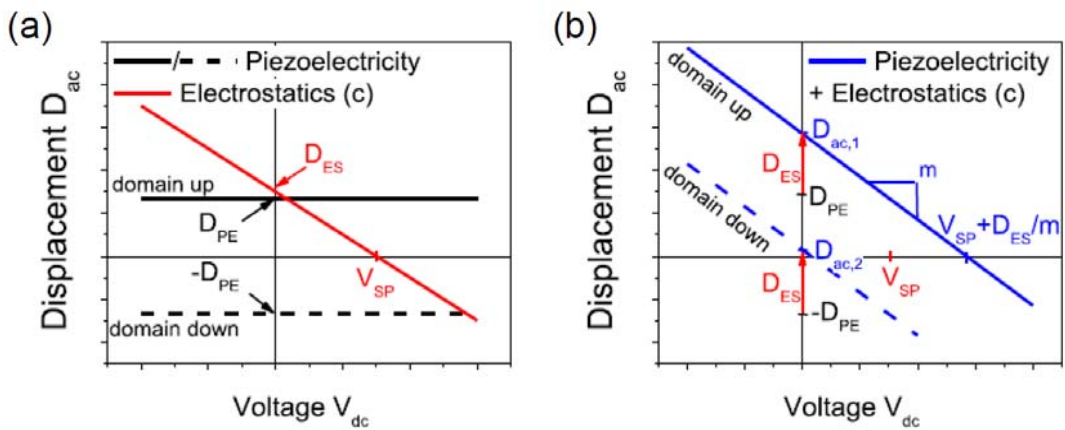

**Fig. 6.** Schematic analysis of the piezoelectric and electrostatic contributions, considered separately in (a) and together in (b). Reproduced with permission from [S.J. Nina Balke, Pu Yu, Ben Carmichael, Sergei V Kalinin and Alexander Tselev, Nanotechnology, 27 (2016) 425707]. © IOP Publishing. Reproduced by permission of IOP Publishing. All right reserved. [82]

The electrostatic effect also has a significant influence on the quantitative PFM measurements. As mentioned above, the electrostatic effect causes an additional surface displacement that is detected by the first-harmonic deflection of the cantilever, as indicated by equation (5). As a result, the electrostatic effect increases the measured PFM response directly, as shown in Fig. 6.[82] The surface displacement by the electrostatic effect ($D_{ES}$) was linearly dependent on the applied dc voltage, and its presence ultimately increased the total PFM response, although the piezoresponse did not show any dependence on the dc voltage. Balke *et al.* found the electrostatic effect to increase the measured PFM response in the periodically poled lithium niobate sample.[82] This implies that the electrostatic effect is thoroughly taken into account in the quantitative PFM measurement. For instance, the effective piezoelectric coefficient measured by the PFM sometimes gives implausibly large value compared to the expected value. In this case, the electrostatic influence must be considered, regardless of whether it contributes to the measured effective piezoelectric coefficient or not.

**2.2. Minimization of the electrostatic effect**

The electrostatic effect is a typical undesired interaction that occurs in PFM measurements, and that can distort the piezoresponse hysteresis loop as well as hinder the quantitative analysis of the PFM response. Because it inherently originates from the electrostatic force between the AFM tip/cantilever and the sample surface, its influence cannot be entirely removed from the PFM response. However, several approaches can minimize the electrostatic effect. In the piezoresponse hysteresis loop measurements, the electrostatic effect induced by the applied dc voltage can be minimized by using the pulsed dc mode instead of the continuous dc mode, as discussed in Fig. 4. However, it is worth mentioning that, since charge injection into the sample surface can occur while in the on-field state, the electrostatic effect can hardly be expected to be entirely eliminated even in the off-field state. Nonetheless, a piezoresponse hysteresis loop measurement in pulsed dc mode can provide more exact information on the ferroelectric properties of material than one in continuous dc mode. Moreover, it is well known that the influence of the electrostatic effect can be successfully minimized using a cantilever with a relatively high spring constant. This fact can be readily understood from equation (5), in which the electrostatic effect is inversely proportional to the cantilever spring constant.[86, 87] In addition, scanning the edge of the sample surface can be a technically effective method for reducing the electrostatic effect by decreasing the interaction area between the AFM cantilever and the sample surface.[80] Apart from these approaches, top electrode on the sample surface can also alleviate electrostatic effect between the AFM tip and sample.[81, 88] In particular, Kim *et al.*, reported a modified PFM setup that uses top electrode and additional microscopic probe needle for applying voltage instead of the conductive AFM tip.[88] In this setup, because a non-conductive AFM tip can be used for the PFM measurement, electrostatic effect can be effectively removed.

## 2.3. Electrochemical strain

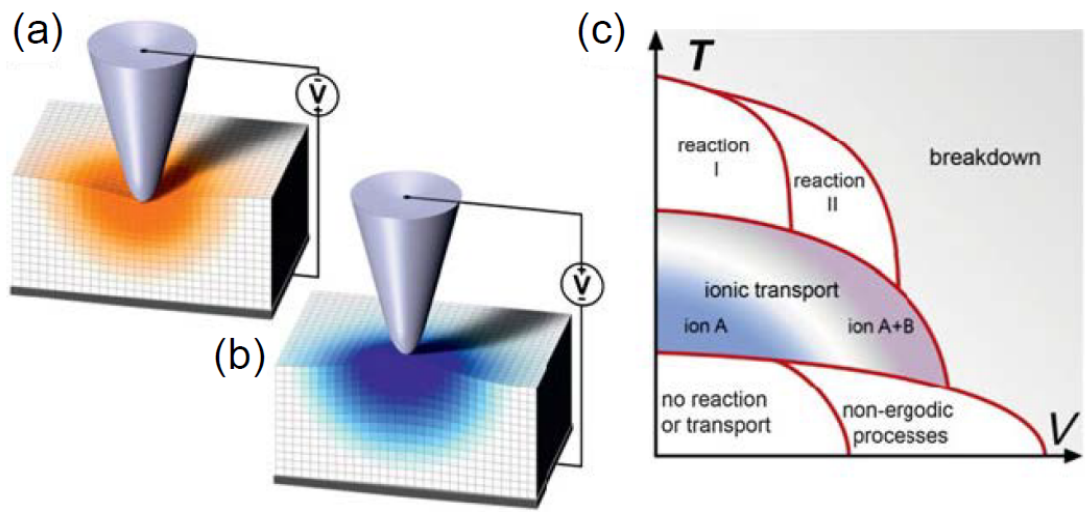

**Fig. 7.** Schematic representation of the electrochemical strain induced by the application of a voltage through an AFM tip in an ionically active material. Accumulation of (a) negative or (b) positive species during the application of a positive or negative tip bias, respectively. Reproduced with permission from [T.M. Arruda, A. Kumar, S.V. Kalinin and S. Jesse,

Nanotechnology, 23 (2012) 325402]. © IOP Publishing. Reproduced by permission of IOP Publishing. All right reserved.[89] (c) Spectrum of possible electrochemical and ionic phenomena, displayed as a function of voltage (*V*) and temperature (*T*), that can be explored by ESM. Reprinted from [S. Jesse, A. Kumar, T.M. Arruda, Y. Kim, S.V. Kalinin and F. Ciucci, MRS Bull, 37 (2012) 651-658] with permission of the Materials Research Society. [90]

The electrochemical (*i.e.*, Vegard) strain can also produce a significant EM response, similar to the piezoresponse in the ionically active materials.[90-92] In general, a surface displacement induced by ionic phenomena (*i.e.*, diffusion and electromigration, and/or electrochemical reactions) is referred to as an electrochemical strain. In the AFM-based approaches, such electrochemical strain can be induced by applying a voltage and probed using a conductive AFM tip, as described schematically in Figs. 7(a) and (b).[89] Specifically, in the ionically active material, an external electric field produced by a voltage applied to the AFM tip produces a local ion redistribution underneath the AFM tip, which in turn causes local volume change, eventually resulting in a displacement of the surface. The surface displacement can be probed by observing the vertical deflection of the AFM cantilever, which could provide local information on the electrochemical properties of material surface. Morozovska and Kumar *et al.*, reported that this ion redistribution as well as irreversible electrochemical reaction by the external electric field can be induced even in the range of kHz and effectively probed by the AFM cantilever.[92, 93] On the basis of this principle, Kalinin's group suggested an AFM-based technique, referred to as electrochemical strain microscopy (ESM).[90] The observed electrochemical strain in the ESM depends on the magnitude and frequency of the applied voltage and temperature, and can thus be expressed as

$$(PR)_{\omega,EC} = \varepsilon(V, \omega, T)V_{ac}\sin(\omega t). \tag{6}$$

where $\varepsilon(V, \omega, T)$ is the effective electrochemical strain constant, expressed as a function of the magnitude $V$ and frequency $\omega$ of the applied voltage and the temperature $T$. Note that $\varepsilon$ is defined as the electrochemical strain coefficient in a similar manner to the piezoelectric coefficient.[94] Despite having significantly different fundamental mechanisms, the electrochemical strain and piezoresponse are eventually detected as an EM response in PFM and ESM, respectively. This fact implies that the EM response produced by electrochemical strain is potentially misinterpreted as that produced by the piezoresponse.[95-97]

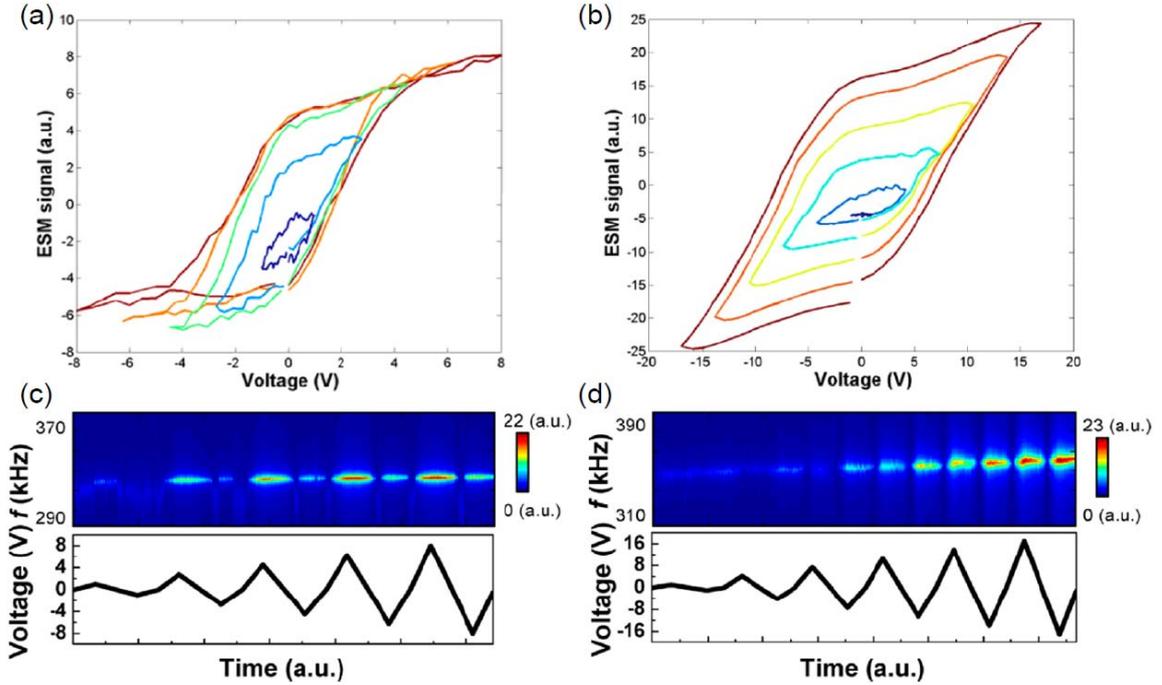

**Fig. 8.** (a, b) Average first-order reversal-curve (FORC)-type ESM loops, drawn over a spatial grid of points. (c,d) (top panels) Single-point BE amplitude spectra as functions of a varying bias voltage (bottom panels) during the FORC acquisition in (a, c) SrTiO$_3$ and (b, d) TiO$_2$ thin films. Reprinted with permission from [Y. Kim, A. N. Morozovska, A. Kumar, S. Jesse, E. A. Eliseev,

F. Alibart, D. Strukov and S. V. Kalinin, ACS nano, 8 (2012) 7026]. Copyright (2012) ACS Publications. [95]

Indeed, electrochemical strain can reportedly generate a ferroelectric-like EM hysteresis loop with respect to the applied triangular voltage waveform, *i.e.*, a butterfly-like shape displaying amplitude hysteresis loop and a 180° phase difference.[95, 97-99] Figure 8 shows that clear ferroelectric-like EM hysteresis loops can be observed in non-ferroelectric $SrTiO_3$ (STO) and $TiO_2$ thin films.[95] Even though STO often displays real ferroelectricity in specific cases, such as the strained state resulting from lattice mismatch,[100-102] $TiO_2$ does not display real ferroelectricity. As a result, the observed hysteric behavior is unlikely to be related with ferroelectricity. The origin of this hysteric behavior can be attributed to the electrochemical strain, depending on the motion of oxygen vacancies driven by the external electric field.[103] Since it is well known that a positively charged oxygen vacancy is mobile in oxide thin films, the applied electric field can redistribute the oxygen vacancies. This in turn produces an electrochemically induced surface displacement, depending on the degree of accumulation and depletion of oxygen vacancies under the AFM tip. By this mechanism, a clear ferroelectric-like EM hysteresis loop can be generated even in non-ferroelectric oxide materials, as shown in Figs. 8(a) and (b). The increase in the hysteresis loop area depends on the applied voltage and clearly indicates that oxygen vacancy motions become more active when a high voltage is applied. The surface displacement then increases accordingly. This behavior was also clearly observed in BE amplitude spectra, as shown in Figs. 8(c) and (d). The gradual evolution of the amplitude signal demonstrates the voltage dependence of the observed hysteresis loop.

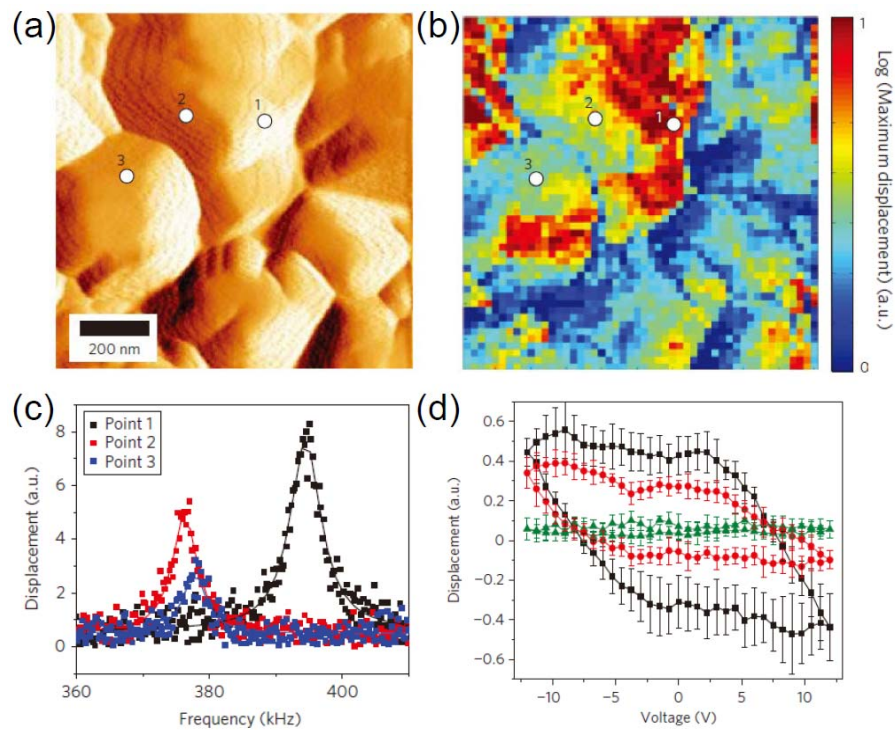

**Fig. 9.** (a) Deflection signal of an area of size 1 μm × 1 μm. (b) Maximum displacement (amplitude) of the contact resonance peak measured in a 50 × 50 point grid in the same area shown in (a). (c) Strain response of the surface to 1 $V_{ac}$ applied to the tip at the three marked locations in (a). (d) Measured electrochemical hysteresis loops at different locations. Reproduced with permission from [N. Balke, S. Jesse, A.N. Morozovska, E. Eliseev, D.W. Chung, Y. Kim, L. Adamczyk, R.E. Garcia, N. Dudney and S.V. Kalinin, Nat. Nanotechnol, 5 (2010) 749-754]. [104]

In addition to conventional oxide systems, ESM studies have reported an electrochemically induced EM hysteresis loop in the lithium-ion-based cathode material $LiCoO_2$, presented in Fig. 9.[104] In an ionic material system, the redistribution of lithium ions occurs primarily when an external electric field is applied. The ionic motion then generates electrochemical strain.[105] Local electrochemical strain clearly differs in grains and grain boundaries, as shown in Fig. 9(b). This observation directly indicates that the degree of redistribution of lithium ions by the electric field was also spatially variable. The measured EM hysteresis loops differed accordingly, as shown in Fig. 9(d). In this case, because the hysteresis loop area represents the degree of redistribution of lithium ions and is directly proportional to the magnitude of the electrochemical strain,[106, 107] it is inferred that the large hysteresis loop can be obtained from the strong redistribution of lithium ions. Whereas the hysteresis loop with a small area does not show clear hysteric behavior, the large loop indicates ferroelectric-like hysteric behavior. Consequently, the electrochemical strain induced by a sufficiently strong redistribution of lithium ions can generate a ferroelectric-like EM hysteresis loop.

Overall, these observations demonstrate that ionically induced electrochemical strain can be one of origins of the EM response and, further, create an EM hysteresis loop analogous to the piezoresponse hysteresis loop, even in non-ferroelectric materials. Accordingly, the hysteresis loop observed by PFM cannot be used as concrete evidence for ferroelectricity. The electrochemical strain must also be considered as a possible origin of the observed hysteresis loop, to establish the presence of ferroelectricity reliably.

## 2.4. Other non-piezoelectric origins

### 2.4.1. Electrostriction and Flexoelectricity

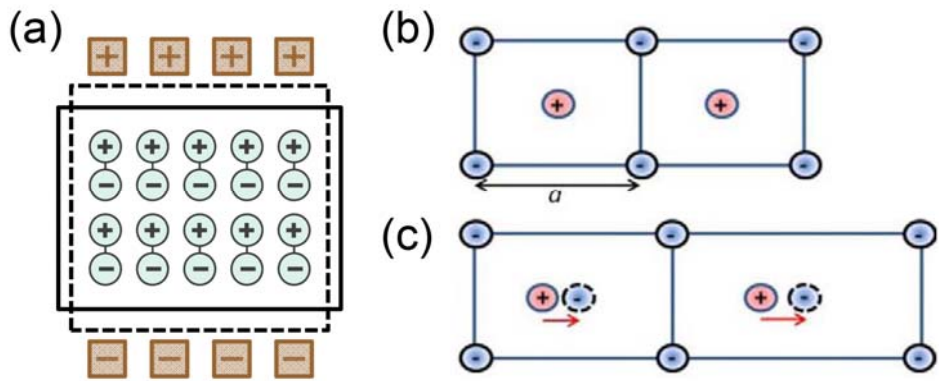

**Fig. 10.** Basic mechanism of the (a) electrostrictive and (b-c) flexoelectric effects. Diagram (b) shows the undeformed stress-free configuration of a portion of a two-dimensional diatomic ionic solid. Diagram (c) shows the deformed configuration, wherein each atom is subjected to an inhomogeneous displacement. Panel (b) and (c) are reprinted with permission from [R. Maranganti and P. Sharma, Phys. Rev. B, 80 (2009) 054109] [108]. Copyright (2009) by The American Physical Society.

Figure 10(a) shows how, under the application of an external electric field, a small shape change can occur in any dielectric material. This effect, showing a quadratic relationship with respect to the applied electric field, is called electrostriction.[110, 111] It is well known that the interaction between the charged dipoles in a material and the external electric field results in electrostriction. Eliseev *et al.* theoretically demonstrated the contribution of electrostriction to the PFM and ESM responses.[111] They showed that this contribution cannot be readily separated from the PFM and ESM responses, and furthermore, it is strongly dependent on the properties of the measured materials. This indicates that electrostriction can also affect the observed PFM response. Nonetheless, it is generally expected that electrostriction may be small compared to the piezoresponse and electrochemical strain.[112]

Flexoelectricity is an interaction between the electrical polarization and the mechanical strain gradient, as described in Figs. 10(b) and (c).[108, 109, 113, 114] Because flexoelectricity is related with the strain gradient, its effect becomes more significant in thin films and other nanostructures through, *e.g.*, lattice mismatch at the interfaces. Flexoelectricity is also known to exist in all materials, regardless of their structural symmetry.[113] This effect reportedly affects the ferroelectric properties of materials, such as their critical thickness, the electrical transport of the domain wall, and the mechanical switching of polarization.[115-117] It consequently also affects the observed PFM response by unpredictably changing the material properties. Nonetheless, its impact on the observed PFM response may be very small compared to those of the other contributions. For instance, Seol *et al.* reported that it is two orders of magnitudes smaller than the electrochemical strain.[112] However, flexoelectricity is still poorly understood and, furthermore, its contribution cannot be clearly discriminated from the ionically induced chemical effect, *e.g.*, the electrochemical strain.[114] Further investigation will be necessary to clarify the contribution of the flexoelectricity to the observed PFM response explicitly.

**2.4.2. Joule heating**

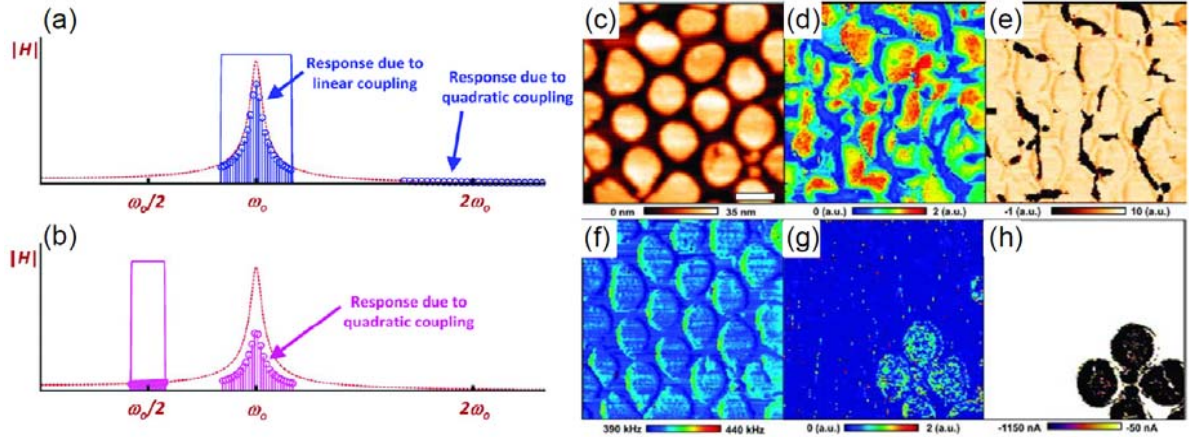

**Fig. 11.** Operational principle of (a) the first $\omega_0$ and (b) the half $\omega_0/2$ harmonic BE-PFM modes. (c) Topography, first-harmonic BE-PFM (d) amplitude and (e) phase, (f) resonance frequency, (g) half-harmonic BE-PFM amplitude, and (h) current maps of BFO nanocapacitors obtained with a bias of -250 mV applied to the conductive AFM tip. The scale bar is 400 nm. Reprinted with permission from [Y. Kim, A. Kumar, A. Tselev, I. I. Kravchenko, H. Han, I. Vrejoiu, W. Lee, D. Hesse, M. Alexe, S. V. Kalinin and S. Jesse, ACS nano, 11 9104 (2011)]. Copyright (2011) ACS Publications.[118]

Joule heating refers to a temperature increase generated by current flow. This can cause an EM response via thermal expansion, which can in turn contribute to the PFM response. In PFM, the surface displacement induced by Joule heating can be expressed as[79, 118]

$$x_J = \beta I^2 R, \qquad (7)$$

where $\beta$, $I_{ac}$, and $R$ are the Joule heat transduction coefficient, the applied current magnitude, and the resistance, respectively. Since $I = I_{ac}\sin(\omega t)$ is applied, the above equation can be

expressed as

$$(PR)_{2\omega,J} = -\beta \frac{RI_{ac}^2}{2} \cos(2\omega t). \tag{8}$$

Furthermore, if the current-voltage characteristics obey Ohm's law, eq. (8) can be rewritten as

$$(PR)_{2\omega,J} = -\beta \frac{V_{ac}^2}{2R} \cos(2\omega t). \tag{9}$$

It is clear that Joule heating contributes to the second-harmonic response, and that its effect can be detected by observing the second-harmonic response of the cantilever deflection during a PFM measurement. On the basis of this relation, the effect of Joule heating on the EM response was explored using the half-harmonic BE (HBE) mode, as shown in Fig. 11.[118] Figures 11(a) and (b) represent the operation principle of the HBE mode. In this mode, the sample is excited by applying an HBE waveform at half the contact-resonance frequency $\omega_0/2$ to enhance the magnitude of the second-harmonic response using the contact-resonance frequency $\omega_0$. As discussed above, since Joule heating contributes to the second-harmonic response, its effect can be observed by this approach. The results obtained for the $BiFeO_3$ (BFO) nanocapacitor are shown in Figs. 11(c-h). Interestingly, the relatively leaky nanocapacitor, as clearly distinguished in the current image of Fig. 11(h), displays a higher HBE-amplitude response. This observation indicates that Joule heating by a significant current flow contributes to the EM response in leaky capacitors. It is noted that, even though the contribution of Joule heating displays a quadratic relationship with respect to the applied voltage, the quadratic relationship can also affect the first-harmonic response for ac and dc voltage sweeps, for instance, the ac voltage sweep used for piezoelectric coefficient measurements.[119-121] As a result, Joule heating can potentially affect quantitative a PFM measurement if the sample is relatively leaky.

Overall, there are potential contributions to the PFM signals by the electrostriction,

flexoelectric effect and Joule heating. The contributions of these non-piezoelectric effects can hardly be separated in the measured PFM signals explicitly due to their concurrent nature. For instance, electrostriction and Joule heating simultaneously contribute to the second harmonic EM response. This indicates that each contribution is hard to be solely determined even in the harmonic measurement. Consequently, it is important to recognize presence and impact of the potential non-piezoelectric effects in the PFM measurement.

**3. Differentiation of the ferroelectric contribution from the EM response**

Because there are several contributions to the EM response, the observed PFM response and ferroelectric-like behavior do not always originate from the piezoresponse of ferro/piezoelectric materials. Thus, PFM measurements do not always provide unambiguous evidence for ferroelectricity. Even though ferroelectricity can be confirmed using macro/microscopic measurement techniques, *e.g.*, a polarization-electric field (P-E) hysteresis loop[122-124] and optical second-harmonic generation (SHG),[125-127] these techniques have limited value for exploring the ferroelectricity of nanostructures or nanoscale ferroelectricity. Thus, recognizing the origins of the EM response is very important for understanding the ferro/piezoelectric properties of measured materials. In order to differentiate a ferroelectric contribution from the EM response, several approaches have been suggested, based on observed variations in the EM response due to experimental conditions.[96, 97, 112] We here introduce several approaches that can be readily implemented using PFM.

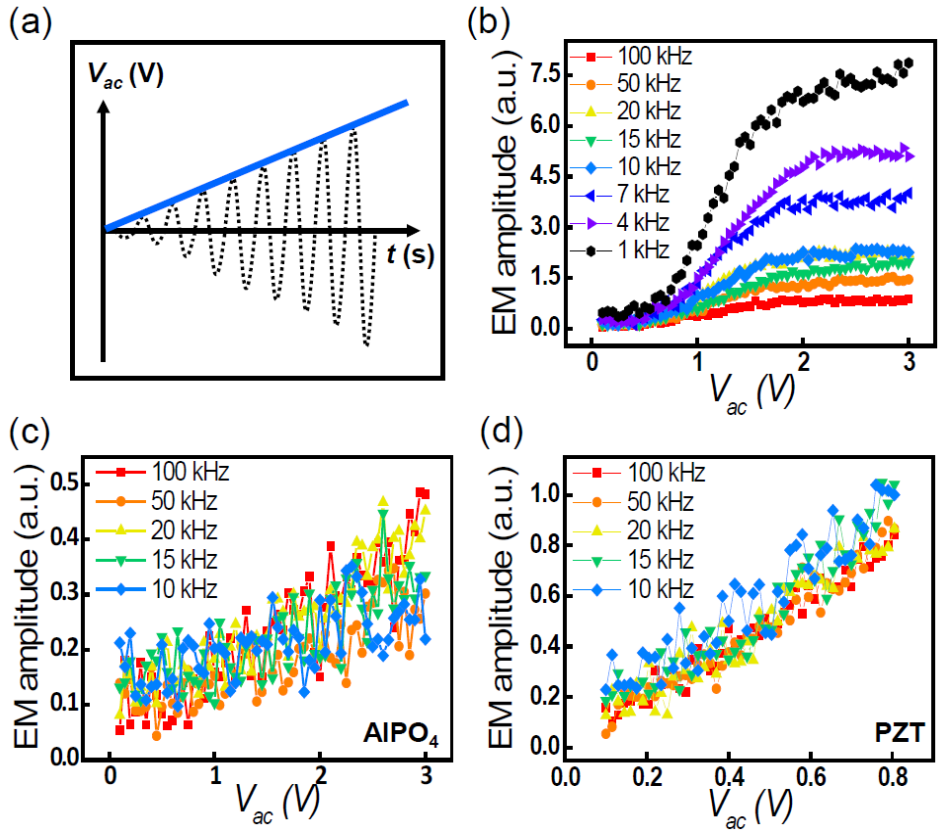

**Fig. 12.** (a) Schematic of the ac amplitude sweep. (b-d) EM amplitudes of (b) the LAGTPO phase of the LICGC, (c) the AlPO$_4$ phase of the LICGC and (d) a PZT thin film, plotted as functions of the ac amplitude. Reproduced with permission from [D. Seol, S. Park, O.V. Varenyk, S. Lee, H.N. Lee, A.N. Morozovska and Y. Kim, Sci. Rep. 6 (2016) 30579]. [112]

It was previously reported that a differentiation of the EM responses induced by the piezoresponse and electrochemical strain was possible by observing the frequency dependence in PZT thin films and lithium-ion conductive glass ceramics (LICGCs) samples.[112] Since LICGC simultaneously exhibits ionic (Li$_x$Al$_x$Ge$_y$Ti$_{2-x-y}$P$_3$O$_{12}$ (LAGTPO)) and piezoelectric phases (AlPO$_4$), it can be a suitable model sample to investigate different EM responses on the same surface.[94] In order to explore the frequency dependence of the EM response, a frequency-dependent amplitude sweep was employed, as schematically depicted in Fig. 12(a). In this approach, the periodic EM response induced by applying a gradually increasing ac voltage was measured over various frequency ranges. Interestingly, the EM response measured on the LAGTPO phase of the LICGCs shows a significant frequency dependence, as shown in Fig. 12(b). In other words, the EM response increases as the frequency of the ac amplitude decreases. This behavior is readily understood, given that the EM response of the LAGTPO phase originates from the electrochemical strain due to the motion of lithium ions, and that the degree of ionic motion is known to increase as the frequency decreases.[92] On the other hand, the EM responses measured on the piezoelectric $AlPO_4$ and the ferroelectric PZT thin film did not change with frequency, as clearly seen in Figs. 12(c) and (d). These results indicate that the EM response induced by the piezoresponse does not exhibit a significant frequency dependence in the measured frequency range. Consequently, the ferroelectric contribution can be differentiated from the EM response on the basis of their different frequency dependences..

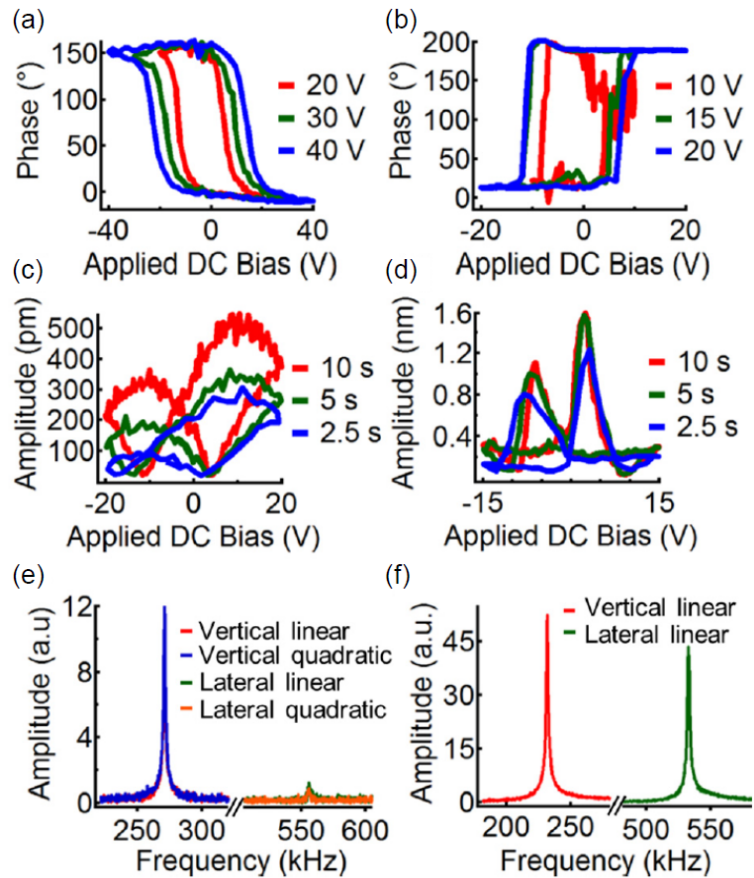

**Fig. 13.** Comparison of spontaneous and induced ionic polarizations, in terms of hysteresis loops of (a) soda-lime glass and (b) PZT in a vertical PFM under different maximum DC voltages; butterfly loops of (c) soda-lime glass and (d) PZT in vertical PFM for different cycling periods; and vertical and lateral PFM response of (e) soda-lime glass and (f) PZT. Reproduced with permission from [Q. N. Chen, Y. Ou, F. Ma and J. Li Appl. Phys. Lett., 104 (2014) 242907]. Copyright 2014, AIP Publishing LLC. [97]

Chen *et al.* reported a difference in the hysteric behaviors between a ferroelectric PZT thin film and ionic soda-lime glass, which can be used to discriminate between the different origins as shown in Fig. 13.[97] The coercive voltage was not clearly defined for the electrochemically induced EM response of the soda-lime glass. Because the application of a higher voltage can increase the degree of ionic motions, the area of the corresponding hysteresis loops also increases with increasing applied maximum voltage. On the other hand, the coercive voltage for the PZT thin film was not dependent on the magnitude of the applied maximum voltage because polarization switching occurs in the vicinity of the coercive voltage regardless of the applied maximum voltage. Furthermore, a longer period in the voltage should result in a greater electrochemical strain, whereas a significant time dependence is not expected for ferroelectric polarization switching. Indeed, different time-dependent hysteric behaviors were clearly observed in Figs. 13(c) and (d). The two origins show significantly different time dependences and hence, a time-dependent hysteresis measurement can effectively discriminate between them. It is noted that time dependence of the ferroelectric polarization switching can be also shown if voltage sweep time is very short as compared to polarization switching time. However, since hysteresis loop measurement in the PFM is generally carried out in the range of a few Hz, time dependence of the ferroelectric polarization switching can be minimal. In addition to these different hysteric behaviors, the electrochemical strain and piezoresponse also differ in their lateral responses. That is, the shear response of the electrochemical strain in soda-lime glass is relatively weaker than the vertical response owing to its amorphous nature. In contrast, the lateral response is comparable to the vertical one in the piezoresponse because there can be a lateral deformation when the surface volume is changed by the converse piezoelectric effect. As presented in Fig. 13(e) and (f), the lateral response of the electrochemical strain was much weaker than the vertical one, whereas the piezoresponses were nearly the same in both cases.

These observations provide an additional effective method to discriminate between the ferroelectric contributions from the EM response.

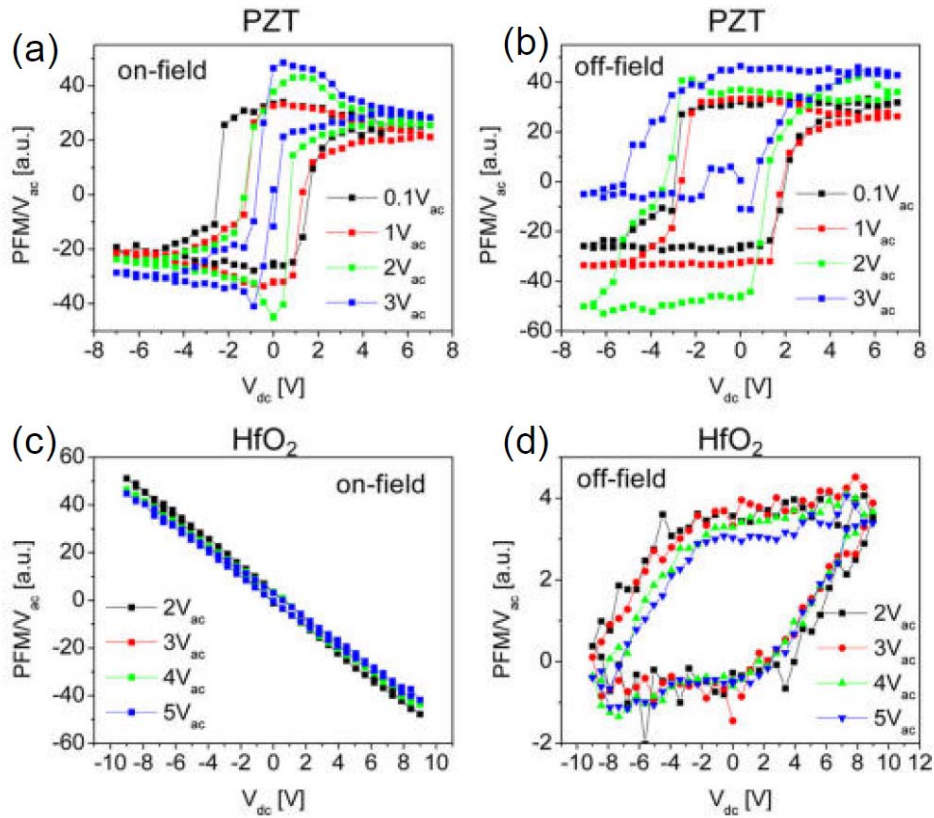

**Fig. 14.** (a) On-field and (b) off-field PFM hysteresis loops plotted as functions of the ac voltage magnitude for PZT. Similarly, (c) and (d) are the corresponding PFM hysteresis loops for $HfO_2$. All measured PFM values are normalized by the ac voltage magnitude. Reprinted with permission from [N. Balke, P. Maksymovych, S. Jesse, A. Herklotz, A. Tselev, C.-B. Eom, I. I. Kravchenko, P. Yu and S. V. Kalinin, ACS nano, 9 (2015) 6484]. Copyright (2015) ACS Publications. [96]

Balke *et al.* recently reported that the ferroelectric contributions can be differentiated on the basis of the shape of the hysteresis loop, depending on the magnitude of the probing ac voltage with respect to the coercive voltages.[96] The shape of the hysteresis loops, based on the piezoresponse, depends on the ac voltage magnitude because the probing ac voltage also contributes to polarization switching when the applied ac voltage exceeds the coercive voltage.[128] Figure 14 plots the obtained on- and off-field hysteresis loops as functions of the ac voltage magnitude in ferroelectric PZT and non-ferroelectric $HfO_2$ thin films. The shape of hysteresis loops based on the piezoresponse clearly depends on the probing ac voltage (Figs. 14(a) and (b)). On the other hand, non-ferroelectric $HfO_2$ showed no significant dependence (Figs. 14(c) and (d)).

In conclusion, several different approaches for differentiating the ferroelectric contribution from the EM response have been discussed in this review. However, despite their effectiveness in the context of the observed results, these methods sometimes fail to clarify the origins of the EM response unambiguously. In this case, a combination of suggested approaches can be one of the solutions. Moreover, utilization of the micro/macroscopic approaches for investigating ferro/piezoelectricity, *e.g.*, P-E hysteresis loop and optical SHG, can be also another solution that underpins observed PFM results.

## 4. Summary

PFM has established itself as a versatile and powerful tool for investigating ferro/piezoelectric phenomena at the nanoscale. In many cases, it provides useful insight into ferro/piezoelectricity. However, undesired non-piezoelectric effects can induce an EM response that resembles the piezoresponse. The resulting contributions can indeed dominate the measured PFM response and, further, mimic ferroelectric behavior. This can in turn result in a serious misinterpretation of PFM measurements. The presence of these non-piezoelectric effects and their impact must therefore be recognized. The electrostatic effect impacts on the shape of the piezoresponse hysteresis loop as well as the quantitative PFM response. Furthermore, a ferroelectric-like EM hysteresis loop can even be induced by the electrochemical strain. In addition, there are potential contributions from electrostriction, flexoelectricity, Joule heating, etc. Even though several AFM based approaches for differentiating the ferroelectric contribution are suggested, it is hard to be expected unambiguous separation between ferro/piezoelectric and non-piezoelectric effects. Consequently, concurrent non-piezoelectric effects are always appropriately considered for understanding intrinsic ferro/piezoelectric properties of a material in the PFM measurement. Ultimately, by taking these effects into account, the applicability of PFM could be extended to other material systems.

**Acknowledgements**

This work was supported by the Basic Science Research Program through the National Research Foundation of Korea (NRF), funded by the Ministry of Science, ICT, & Future Planning (NRF-2014R1A1A1008061 and NRF-2014R1A4A1008474).